\documentclass[conference]{IEEEtran}
\IEEEoverridecommandlockouts

\usepackage{tikz}
\usepackage{textcomp}

\newcommand\copyrighttext{%
  \footnotesize \textcopyright 2022 IEEE. Personal use of this material is permitted.
  Permission from IEEE must be obtained for all other uses, in any current or future
  media, including reprinting/republishing this material for advertising or promotional
  purposes, creating new collective works, for resale or redistribution to servers or
  lists, or reuse of any copyrighted component of this work in other works.
  ISBN: 978-1-6654-0627-7}
\newcommand\copyrightnotice{%
\begin{tikzpicture}[remember picture,overlay]
\node[anchor=south,yshift=10pt] at (current page.south) {\fbox{\parbox{\dimexpr\textwidth-\fboxsep-\fboxrule\relax}{\copyrighttext}}};
\end{tikzpicture}%
}

\usepackage{cite}
\usepackage{amsmath,amssymb,amsfonts}
\usepackage{algorithmic}
\usepackage{graphicx}
\usepackage{textcomp}
\usepackage{xcolor}
\def\BibTeX{{\rm B\kern-.05em{\sc i\kern-.025em b}\kern-.08em
    T\kern-.1667em\lower.7ex\hbox{E}\kern-.125emX}}
    
\usepackage{acronym}
\newacro{nic}[NIC]{Network Interface Controller}
\newacro{irq}[IRQ]{Interrupt}
\newacro{dma}[DMA]{Direct Memory Access}
\newacro{cpu}[CPU]{Central Processing Unit}
\newacro{isr}[ISR]{Interrupt Service Routine}
\newacro{ist}[IST]{Interrupt Service Task}
\newacro{tcm}[TCM]{Tightly Coupled Memory}
\newacro{uart}[UART]{Universal Asynchronous Receiver/Transmitter}
\newacro{mmu}[MMU]{Memory Management Unit}
\newacro{nop}[NOP]{No Operation}
\newacro{bd}[BD]{Buffer Descriptor}
\newacro{plc}[PLC]{Programmable Logic Controller}
\newacro{mmio}[MMIO]{Memory-Mapped Input/Output}
\newacro{rx}[RX]{Receive}

\newacro{os}[OS]{Operating System}
\newacro{rtos}[RTOS]{Real-Time Operating System}
\newacro{posix}[POSIX]{Portable Operating System Interface}
\newacro{rms}[RMS]{Rate Monotonic Scheduling}
\newacro{wcet}[WCET]{Worst Case Execution Time}
\newacro{tcb}[TCB]{Thread Control Block}
\newacro{ipc}[IPC]{Inter-process Communication}
\newacro{api}[API]{Application Programming Interface}
\newacro{fifo}[FIFO]{First In First Out}

\newacro{osi}[OSI]{Open Systems Interconnection}
\newacro{phy}[PHY]{Physical Layer}
\newacro{mac}[MAC]{Medium Access Control}
\newacro{ip}[IP]{Internet Protocol}
\newacro{arp}[ARP]{Address Resolution Protocol}
\newacro{icmp}[ICMP]{Internet Control Message Protocol}
\newacro{udp}[UDP]{User Datagram Protocol}
\newacro{tcp}[TCP]{Transport Control Protocol}
\newacro{sctp}[SCTP]{Stream Control Transmission Protocol}
\newacro{quic}[QUIC]{Quick \acs{udp} Internet Connections}
\newacro{dns}[DNS]{Domain Name System}
\newacro{dhcp}[DHCP]{Dynamic Host Configuration Protocol}
\newacro{mtu}[MTU]{Minimum Transmission Unit}

\newacro{csv}[CSV]{Comma Seperated Values}

\newacro{pc}[PC]{Personal Computer}
\newacro{iot}[IoT]{Internet of Things}
\newacro{dos}[DoS]{Denial of Service}
\newacro{qos}[QoS]{Quality of Service}
\newacro{tsn}[TSN]{Time Sensitive Networking}

\newacro{bsd}[BSD]{Berkely Software Distribution}
\newacro{lwip}[lwIP]{Lightweight IP}
\newacro{lrp}[LRP]{Lazy Receiver Processing}

\newacro{hp}[HP]{High Priority}
\newacro{lp}[LP]{Low Priority}
\newacro{mp}[MP]{Medium Priority}
\newcommand{\parencite}[2][]{\ifthenelse{\equal{#1}{}}{\cite{#2}}{\cite[p. #1]{#2}}}
\usepackage{csquotes}
\usepackage{pgfplots}
\usepackage[para]{footmisc}
\usepackage{url}

\begin{document}

\title{Differentiating Network Flows for\\Priority-Aware Scheduling of Incoming\\Packets in Real-Time IoT Systems}

\author{
    \IEEEauthorblockN{Christoph Blumschein\IEEEauthorrefmark{1}, Ilja Behnke\IEEEauthorrefmark{1}, Lauritz Thamsen\IEEEauthorrefmark{2}, and Odej Kao\IEEEauthorrefmark{1}}
    \IEEEauthorblockA{Corresponding: \textit{i.behnke@tu-berlin.de}}
    \IEEEauthorblockA{\IEEEauthorrefmark{1}Technische Universität Berlin}
    \IEEEauthorblockA{\IEEEauthorrefmark{2}University of Glasgow}
}

\maketitle

\copyrightnotice

\begin{abstract}
When \acs{ip}-packet processing is unconditionally carried out on behalf of an \acl{os} kernel thread, processing systems can experience overload in high incoming traffic scenarios.
This is especially worrying for embedded real-time devices controlling their physical environment in industrial \acs{iot} scenarios and automotive systems.

We propose an embedded real-time aware IP stack adaption with an early demultiplexing scheme for incoming packets and subsequent per-flow aperiodic scheduling. %
By instrumenting existing embedded \acs{ip} stacks, rigid prioritization with minimal latency is deployed without the need of further task resources.
Simple mitigation techniques can be applied to individual flows, causing hardly measurable overhead while at the same time protecting the system from overload conditions.
Our \acs{ip} stack adaption is able to reduce the low-priority packet processing time by over 86\% compared to an unmodified stack. %
The network subsystem can thereby remain active at a 7x higher general traffic load before disabling the receive \acs{irq} as a last resort to assure deadlines.
\end{abstract}

\begin{IEEEkeywords}
embedded systems, real-time operating systems, embedded IP stack, internet of things, cyber-physical systems
\end{IEEEkeywords}

\graphicspath{{Figures/}{./}} %

\section{Introduction}
\label{sec:introduction}

In order to react timely to incoming packets, running processes are interrupted by the \ac{nic}.
The resulting processing delays are acceptable in traditional \ac{ip} interconnected devices, i.e. traditional PCs, mobile end-user devices and server systems, as best-effort execution guarantees are satisfactory in these domains.
In contrast, the design of embedded real-time devices needs to take hardware cost, energy efficiency and robustness strongly into account, while computing power is comparably low. At the same time their role as controlling units in a cyber-physical system demands predictable and limited execution times~\cite{alcacer2019scanning}.

It has been demonstrated that an unchanged embedded \ac{ip} subsystem can have disastrous effects on the timing properties of real-time tasks \parencite{behnke_interrupting_2021, niedermaier_you_2018, bender2021pieres}.
With high incoming traffic, almost arbitrarily high delays can be induced, up to the point where the whole system is busy processing \ac{ip} packets. %
Therefore, connecting critical embedded systems to \ac{ip} networks puts real-time properties at risk.
Yet recently, summarized as the \ac{iot} \parencite{jazdi2014cyber}, embedded devices are increasingly getting connected to \ac{ip} networks for the purpose of remote control, reporting measurement data, software maintenance and diagnostics.
As this trend will further expose embedded devices to heterogeneous packet flows in indeterministic \ac{ip}-networks, it gets more important to deal with potentially real-time jeopardizing network-generated interrupts. In order to preserve robust timing behavior in the age of the \ac{iot}, each connected system should be guarded against technical network faults as well as intentional flooding attacks. Addressing this problem will, thus, be necessary for the future deployment of the \ac{iot}.

In contrast to other sources of I/O interrupts, incoming packet events are indeterministic as the developer of an embedded system cannot control the inter-arrival time of network packets. 
However, guarding against overload scenarios via generic rate limitations entails a reduced connection quality and affects real-time properties.
Dedicated hardware solutions are another way this problem can be addressed. These range from dedicating another processor core for networking matters \parencite{realtimehypervisormobileiot} to sophisticated \ac{nic} offloading \parencite{nics, behnke2022priority} and \enquote{Smart \acs{nic}s} equipped with their own fully featured processing system \parencite{smartnics, humphries2019mind}. Moreover, advanced interrupt controllers also address priority space unification~\cite{gomes_task-aware_2015}. Though, as off-the-shelf availability and low cost are central requirements in many IoT environments, custom hardware changes are unlikely to be established. 
Furthermore, best-effort IP networks with wireless links that realize specialized real-time networking technologies (such as \ac{tsn} as proposed in \parencite{8412458}) are still an active topic in research~\cite{bruckner_introduction_2019}.%

In this paper we propose a packet receive architecture for typical embedded \acp{rtos} and \acs{ip}-stacks that facilitates the deployment of IoT hardware in real-time scenarios without the necessity of specialized hardware. More formally, our architecture aims to combine the following properties:
\begin{itemize}
    \item Protection against network-induced system overloads, facilitating real-time systems.
    \item Optimal processing latency for well-behaved \acl{hp} flows.
    \item Best-effort performance for \acl{lp} flows.
\end{itemize}

The paper also introduces a prototypical implementation modifying a popular network stack and presents a set of experiments that evaluate basic performance properties.%

\section{Background}
\label{sec:background}
To better understand the changes made to the embedded IP stack, this section outlines timing relevant aspects of the receive path of network packets and aperiodic scheduling. 

\subsection{Receive Path in \acs{ip} Networking}
\label{sec:rx_path}

\begin{figure}[h]
    \centering
	\includegraphics[width=\columnwidth]{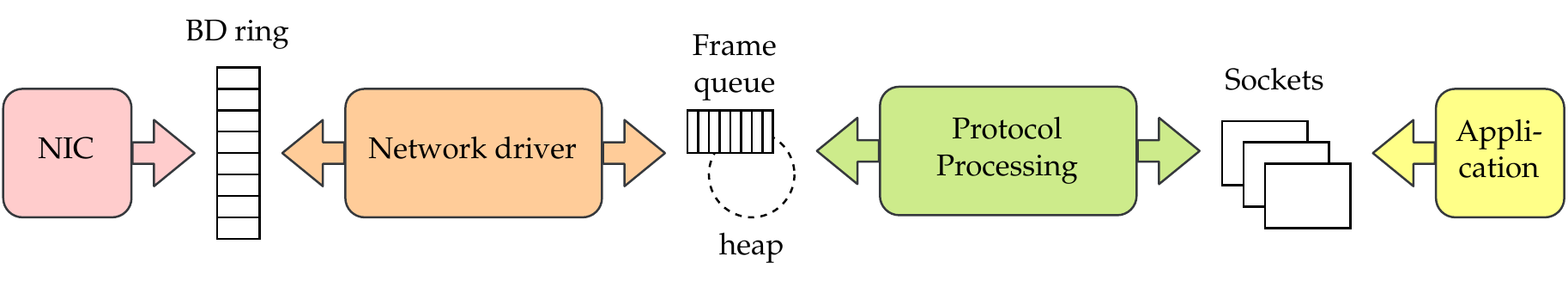}
	\caption{\textbf{\acs{rx}-Path:} Packets are handled by OS data structures, the network driver, and the networking stack implementation before they can be accessed by the receiving application.}
	\label{fig:rx_path_overview}
\end{figure}

At a high level, the \ac{rx} path is organized into subsequently executed stages as seen in Figure \ref{fig:rx_path_overview}.
Upon packet reception, the \ac{nic} transfers the packet content to a previously prepared memory location via \ac{dma}, marks the corresponding \ac{bd} entry and triggers an interrupt.
The network driver, handling the interrupt, acknowledges the \ac{dma}-operation and exchanges the received frame buffer with a newly allocated one.
From here, protocol processing can commence disregarding the already finished MAC-layer operations.
At the end of protocol processing, the set of bound sockets is checked and possibly waiting applications are notified.

\subsection{\acs{irq} Scheduling}
\label{sec:irq}

Hardware \acp{irq}, such as those triggered by incoming network packets, introduce some challenges to scheduling as they might take over CPU resources at any time.
Yet, when systematically tamed to known minimum inter-arrival times and \acp{wcet}, an integration into the considerations of a schedulable task set becomes possible.
More problematic is the triggered execution of \acp{isr} in an elevated \acs{irq}-context.
It may be either completely uninterruptible itself, or only by another higher priority \acs{irq}-source (\enquote{interrupt nesting}).
Hence, \ac{irq} and task priorities form two different priority spaces.

To minimize the worst-case latency incurred by priority inversion situations between interrupts and high-priority tasks, a widespread programming best practice is to reduce the work done in an \ac{isr} to a minimum, only unblocking a deferred \ac{ist} that then does the actual processing.
This compromises on interrupt handling performance for better scheduler control.
Therefore it is often weighed by the driver developer how much additional latency is acceptable until an \ac{isr}/\ac{ist} split is introduced.

\subsection{Aperiodic Scheduling}
\label{sec:aperiodic}
Received network packets generate workload that can be characterized as an aperiodic task inside the \acs{rtos}. 
One approach to integrate aperiodic tasks into fixed-priority scheduling uses so-called server tasks~\cite{sprunt1989aperiodic}.
To the scheduler these behave as ordinary prioritized tasks.
Opposed to other tasks they have no indivdual objective.
Instead, they use their budget to serve the execution of aperiodic jobs.
Due to their limited budget in each period they can be easily included into scheduling considerations.

A very simple yet effective aperiodic server is the deferrable server~\cite{strosnider1995deferrable}, which we also utilize in our real-time receive architecture.
It has a limited CPU-time budget to serve aperiodic events.
When the budget is depleted, it pauses execution until the next renewal.
At the end of each period, it's budget gets restored to the initial amount.
A big advantage is the ease of mechanism and therefore of an implementation for that server scheme.
Yet, the bandwidth preservation property comes at the cost of impairing a higher worst-case processing demand.

Let $p$ be the server period and $e$ it's execution budget for a period.
There may arrive jobs just before the end of a period consuming the whole capacity $e$ of the server for this period.
With the begin of the next period and the consequently budget replenishment, another time $e$ of jobs may be serviced.
So in the worst case, we need to expect one extra execution budget. Thus, the highest possible demand $d(\Delta)$ inside an arbitrary interval can be indicated as
\[ d(\Delta) = e \cdot \left(\left\lceil \frac{\Delta}{p} \right\rceil + 1\right) \]
Note that at least for a small period $p$, this expression still approaches the theoretical server optimum $\frac{e}{p}$.

\subsection{Design Considerations for Embedded \acs{ip}-subsystems}
\label{sec:embedded_network_subsystem_design}

When designing an embedded \acs{ip} network stack, low memory usage is one of the main priorities.
To minimize per-thread data structures like \acp{tcb} and callstack memory, the involved processing is often done in a monolithic \ac{ip}-stack. %
Furthermore, data structures e.g. for mapping port numbers to receiver sockets are typically implemented using a simple list, making the mapping require a primitive iteration instead of more sophisticated lookup mechanism like hash- or order-based ones.
While yielding a much worse time complexity, the small constant factor renders this sufficient for scenarios with relatively few entries, maybe even performing better.
It can be argued that the data instance's size is in full control of the system designer, which still enables the fine worst-case timing analysis required for real-time guarantees.

\section{Related Work}
\label{sec:related_work}
This section presents related work on embedded IP stacks and the problem of network-generated \ac{irq} floods.

Different network stack implementations vary in their set of features. The \textsc{lw}IP network stack is widely spread among embedded applications and sets its focus on memory efficiency~\cite{lwip}. %
Internally, \textsc{lw}IP does not represent a complete data frame but stores a subset of data in \texttt{pbuf} structures. Other stack implementations like FreeRTOS+TCP\footnote{\url{https://www.freertos.org/FreeRTOS-Plus/FreeRTOS_Plus_TCP}} offer the complete frame and stick to the Berkeley sockets API while being thread safe. %

The time-predictable IP stack tpIP~\cite{schoeberl_tpip_2018} addresses the challenge of real-time communication in cyber-physical systems. To enable timing predictability and \acs{wcet} analysis the proposed stack uses polling functions in the socket API with non-blocking read and write operations. While focusing on timing analysis and predictability, no measures are taken towards processing performance, interrupt scheduling or the issue of traffic overloads.

The priority inverting impact of interrupts in real-time systems has been identified and tackled by Amiri et. al. by employing priority inheritance protocols for interrupt service threads~\cite{amiri2015predictable}. This approach however only works for the schedulable part of interrupt handling of device drivers.

Strategies presented in~\cite{packetoverloadmitigation} deal with the detection and mitigation of network packet overloads in real-time systems.
The \emph{Burst Mitigation} approach, limits the amount of \acs{irq}s that may get processed in a time slice, effectively applying a deferrable server scheduling scheme which considers each \acs{irq} a standard-sized job.
While the work does not consider differentiating mitigation measures over different packet flows, the evaluation already hints the practicality of simple mitigation techniques that can be used beneficially in our approach.

Seeking an alternative to the \ac{bsd} TCP/\ac{ip}-stack, Druschel et al. proposed the concept of \ac{lrp}~\cite{druschel1996lazy}.
By introducing a \emph{Early Demultiplexing} stage, where the incoming packets are classified to flows that correspond to the targeted receiver process, they try to improve the thoughput performance, stability and fairness at high incoming network traffic load in server systems.

Building atop the idea of \ac{lrp}, Lee et al. investigated on reducing the impact of \ac{lp}-packets on the real-time behavior of a network-independent task by introducing port-based prioritization of protocol processing \parencite{lee_interrupt_2010, lee_priority-based_2015}.
However their implementation is restricted by the inappropriate scheduling behavior of the \texttt{softirq}-handler in linux, which is not preemptable even by the most critical processes and gets rescheduled in similar way as polling, adding unnecessary high network latency once packets aren't processed eagerly anymore.
Moreover, their work only considers \ac{udp} packets.
Finally, when considering overloading scenarios a mere flow differentiation and prioritization is not sufficient for protecting execution guarantees, since packets may also arrive at a highly prioritized task's port in high quantities.

In order to facilitate evolution of transport protocols, Honda et al. made a point for user-level stacks \parencite{honda2014userlevelstacks}.
Reducing the \ac{os} responsibilities to managing \ac{nic}-sharing and packet multiplexing, these can be referenced as a library by each application independently.
While the code size overhead can be avoided by using shared libraries, the main challenge is enabling elegant and efficient multiplexing and reducing the overhead incurred by user-/kernelspace transitions.

\section{Approach}
\label{sec:approach}

We consider an embedded networking stack running on a \ac{rtos} with a simple fixed-priority scheduler.
Therein a driver controls \acs{dma} transfers, establishes cache coherency and passes packet buffers to a networking task by means of a queue.

\begin{figure}[b]
    \centering
    \includegraphics[scale=0.57]{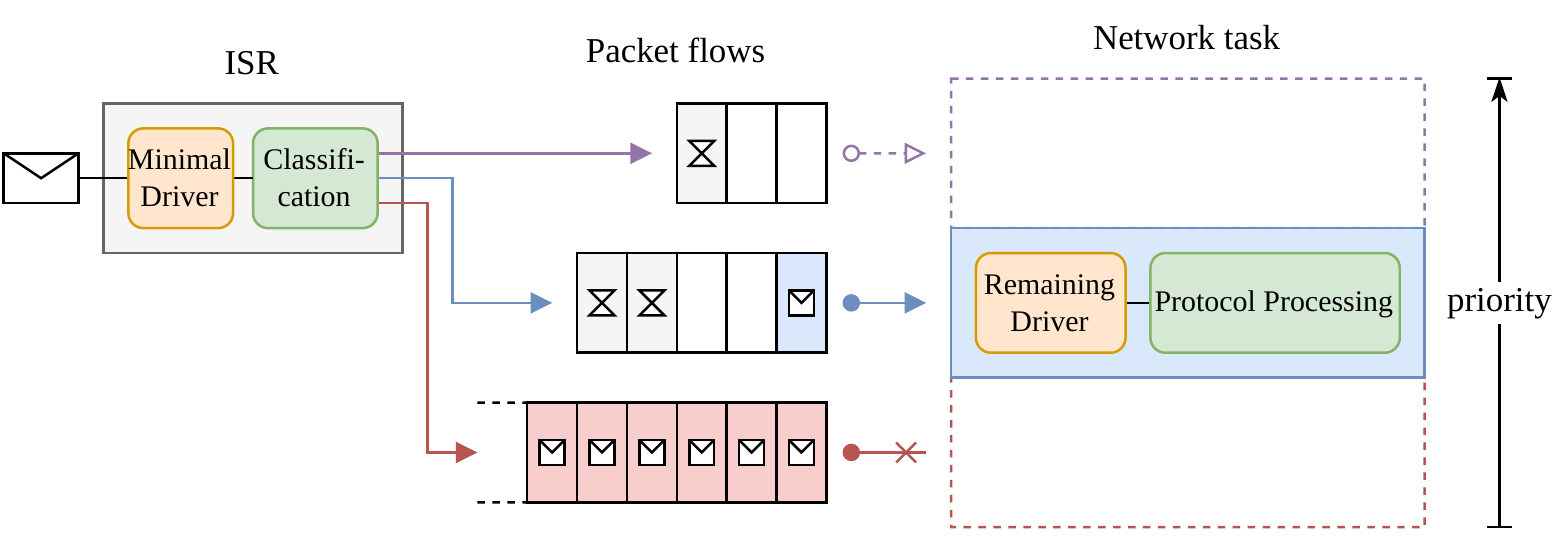}
    \caption{\textbf{Architecture overview:} 
    We classify packets early and enqueue them by their flow, with individual periodic capacity restrictions applied. Then we schedule further processing, including the deferrable driver activity, in a network task with varying priority.}
    \label{fig:approach_overview}
\end{figure}

We then design our architecture around a data structure of differentiated flow queues, which replaces the simple queue.
Each flow defines a priority, limitation capacity and period, to affect the further processing of its packets.
For the prioritization of processing, we add a priority manipulation mechanism to the protocol processing task.
In order to gain a maximal advantage from scheduling the subsequent processing stage, we modify the driver to do only the necessary work of classifying incoming packets to flows by their header entries.
The remaining activity is then executed on packet retrieval by the scheduled protocol processing task (see Figure~\ref{fig:approach_overview}).

Therefore, our proposed architecture combines these three concepts:

\begin{enumerate}
    \item \emph{Soft Early Demultiplexing} into receiver-centric flows.
    \item \emph{Prioritized Protocol Handling} based on these flows.
    \item \emph{Rate Limitation} applied per flow as well as overall.
\end{enumerate}

While all of them are not novel by themselves, we argue that only in this combination they exhibit properties making for a viable solution to the discussed problem:
\begin{itemize}
    \item \emph{Early Demultiplexing} is necessary for differentiating flows on an End-to-End basis, without reliance on network \ac{qos}.
    \item Proper \emph{prioritization} facilitates best-effort communication processes that utilize background resources on the same system.
    \item \emph{Rate limitation} as a last resort protects the system from being vulnerable to unexpectedly high traffic in \ac{hp} flows.
\end{itemize}

In the following subsections we first introduce the concept of each of the three basic building blocks of our architecture and discuss relevant implementation aspects.

\subsection{Soft Early Demultiplexing}

In order to minimize the effort spent until after classification, we employ Early Demultiplexing \cite{druschel1996lazy}.
By peeking into some key header entries, a packet is assigned to its eventual receiver process.

The benefit of demultiplexing performed in software depends heavily on the amount of work that can be saved by mere demultiplexing compared to full protocol processing.
Since the packet scheduling in our architecture can only influence the processing that follows after Early Demultiplexing, the achievable degree of partial network liveness in overload scenarios depends on its quick execution.

Starting from the existing driver receive path, depicted in Figure \ref{fig:driver_buffer_flow}, we introduce two changes:

\begin{figure}[t]
    \centering
    \includegraphics[scale=0.6]{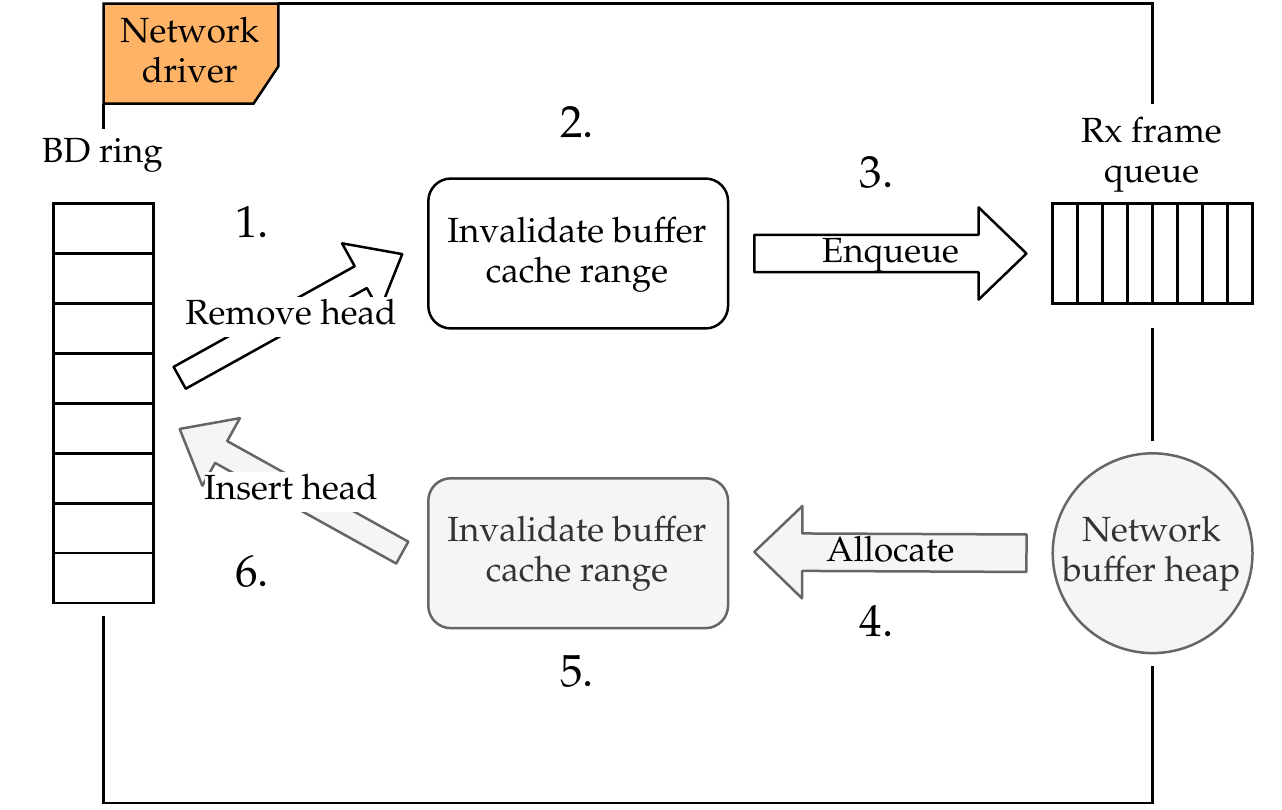}
    \caption[Receive activity in the original driver]{\textbf{Receive activity in the original driver:} 
    Once an Interrupt occurs, the driver code moves a packet buffer from the \acs{bd} ring to a simple queue, and fills the vacant position with a newly allocated one. Due to cache coherency requirements of the memory system, the buffer's caches have to be invalidated for both the retrieved and the replacement buffer.}
    \label{fig:driver_buffer_flow}
    
\end{figure}

\subsubsection{Packet Classification}

The classification differentiates incoming packets into flows defined by the protocols \acs{arp}, \acs{icmp}, \acs{tcp} and \acs{udp}.
While the former two form a single flow, the latter are further differentiated by local port in order to respect the receiver task association.

Depending on the used network stack, the lookup from the port to a flow may either be performed using the existent network stack's list of bound socket control blocks, or else require an additional data structure managed by the driver.
In our prototype based on FreeRTOS+TCP, the socket managing code in the original network stack can easily be locked in a critical section, leaving the \acs{isr} safe to access it.

If a scenario requires anticipating a large number of bound sockets, a sophisticated data structure with better complexity should be employed.
However, with only a few sockets bound at any particular point in time, a linear linked list lookup as found in typical embedded network stacks suffices.

\begin{figure}
    \centering
    \includegraphics[scale=0.5]{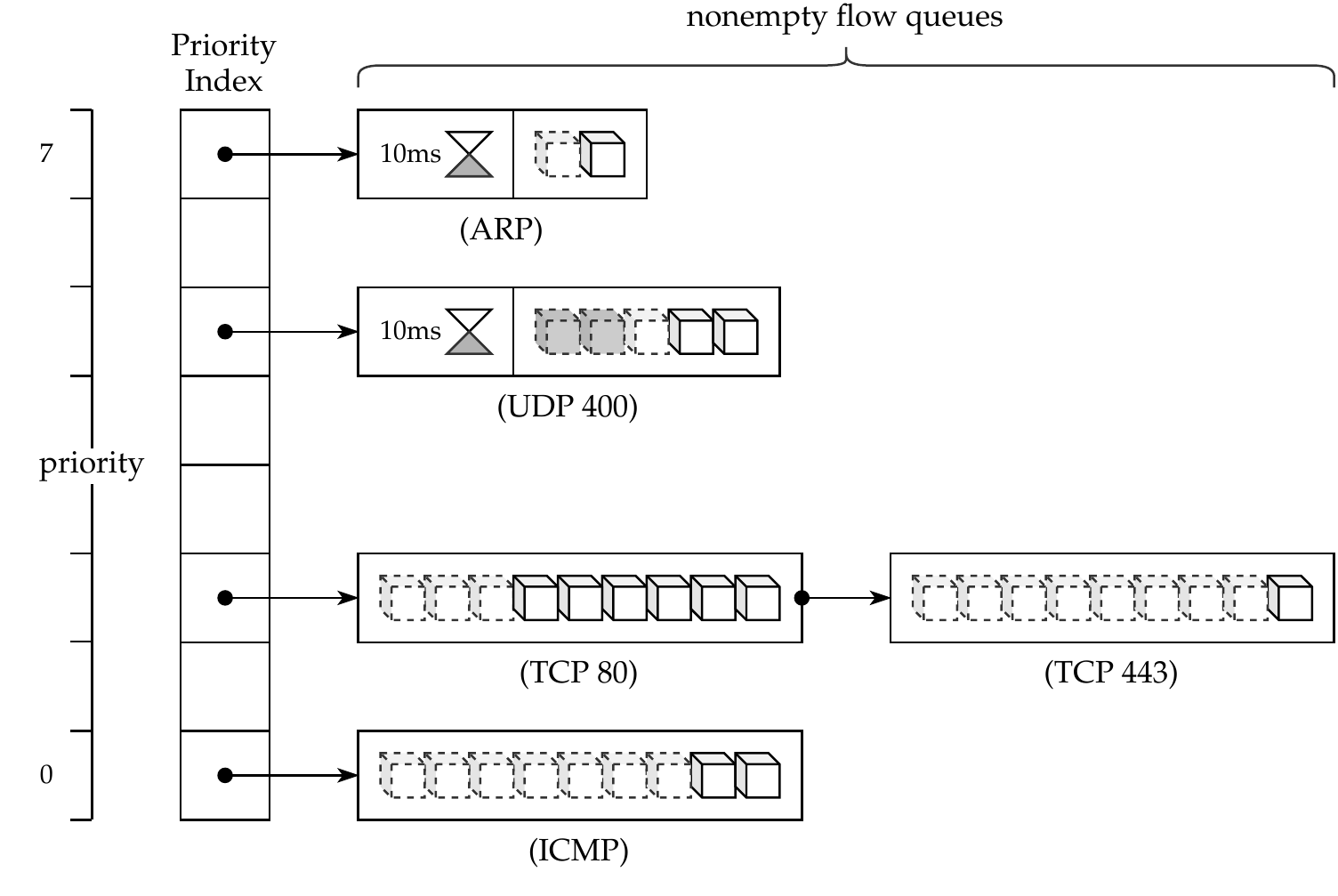}
	\caption[Differentiated flow queues]{\textbf{Differentiated flow queues:} 
	Between reception by the driver and further driver and protocol processing, packets get stored in a queue according to their identified flow. These queues are organized by the flow's priority, to facilitate fast retrieval of the highest/lowest prioritized packet.}
	\label{fig:pdest_depq}
\end{figure}

Instead of enqueueing every received packet to the same \acs{rx} frame queue, each packet is inserted into a specific queue according to the result of the classification.
Because the packets do not necessarily get processed in bounded time, the network subsystem might experience buffer starvation.
To avoid this, packets of low priority are recycled when buffer memory reaches its limit.
To this end, the differentiated flow queues are organized in a priority queue structure (see Figure~\ref{fig:pdest_depq}).
The priority of a flow is defined by its respective receiver task, and the overall priority space is equal to the one used by the \ac{rtos} task scheduler.
Similar to how a fixed priority task scheduler identifies a highest priority task, a lowest priority flow queue can be accessed in constant time, facilitating efficient packet buffer revocation.

\subsubsection{Lazy Cache Invalidation}

On embedded systems that feature CPU-caches, the commonly cache incoherent \ac{dma} introduces a significant cost with the obligation to invalidate the transferred buffer's cache lines.
In our case, the memory architecture requires the network driver to invalidate buffer cache lines prior to and after the processing by the \ac{nic}'s \ac{dma} engine, as depicted in Figure \ref{fig:driver_buffer_flow}.
As cache management noticeably prolongs the execution time of Early Demuxing, we incorporate a lazy cache coherency establishment scheme into the driver.

The driver is therefore split into two halves, as depicted in Figure \ref{fig:my_driver_buffer_flow_with_recycling}:

\begin{figure}[b]
    \centering
    \includegraphics[width=\columnwidth]{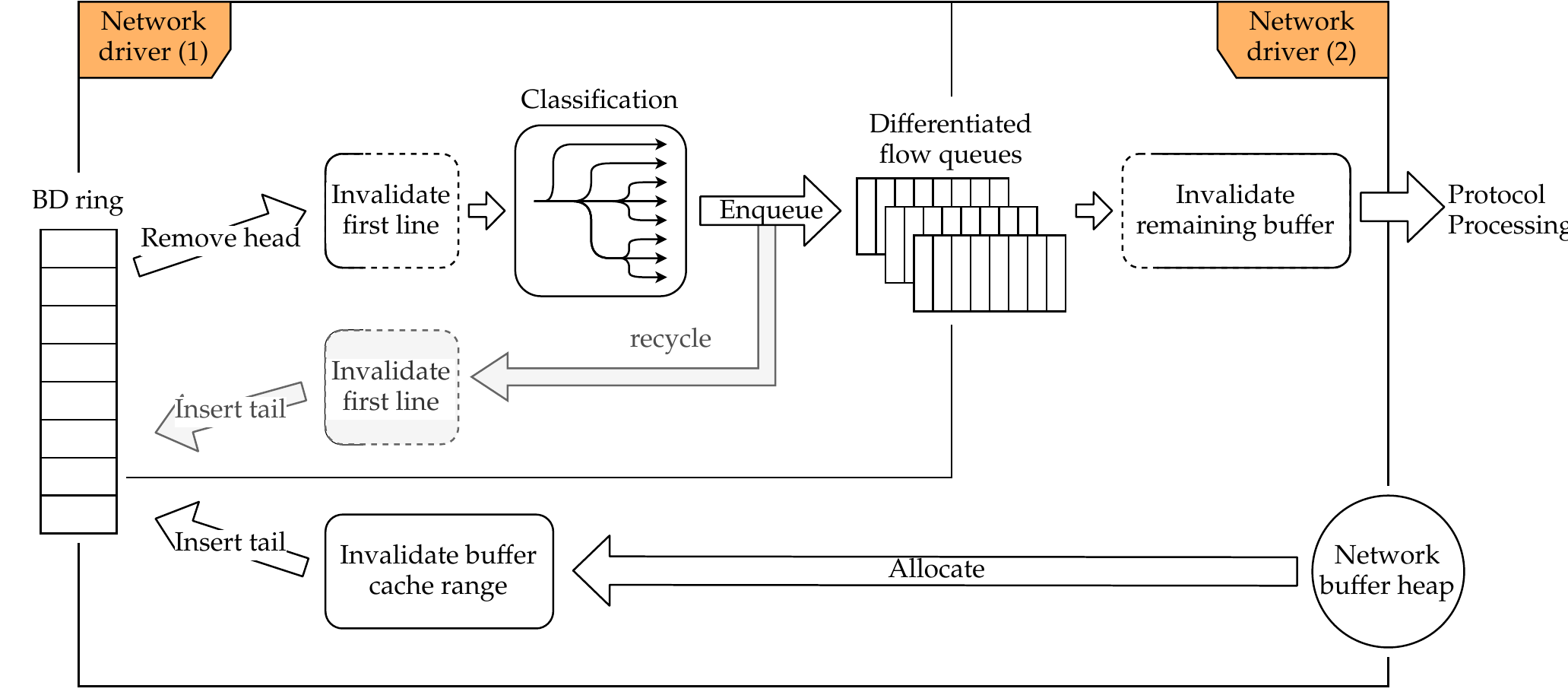}
    \caption{\textbf{Receive driver activity in our approach: }
    The driver is cut in two halves. In the first half, a minimal effort is taken to classify each packet into a flow. As part of the scheduled subsequent protocol processing, the second half establishes cache coherency and refills the \acs{bd} ring once the packet is needed.}
    \label{fig:my_driver_buffer_flow_with_recycling}
\end{figure}

\begin{enumerate}
    \item An immediately processed layer, executed as part of the \ac{isr}, classifies and enqueues packets.
    \item A schedulable layer, executed in the network task according to the packet's priority, establishes full cache coherency of received packets and prepares fresh replacement buffers.
\end{enumerate}

Prior to classification, only the first cache lines of the packet buffer containing the relevant header fields are invalidated.
Once the packet is chosen to be processed further, the remaining part is invalidated and a fresh packet buffer prepared to replace the current buffer in the \ac{dma}'s \ac{bd} ring.
This implies that as the differentiated flow queues fill up with packets, the \ac{bd} ring empties, forming a closed pool of packets shared by the \ac{bd} ring and the differentiated flow queues.
To prevent starvation of the \ac{bd}-ring caused by \acs{lp}-packets stuck in the differentiated flow queues, the immediate driver part recycles lowest priority packets once the \ac{bd}-ring hits a critical threshold (e.g. $\frac{1}{2}$).
This can be carried out with little cost, since only the accessed header cache lines have to be invalidated again.

\begin{figure}
    \centering
    \includegraphics[scale=0.6]{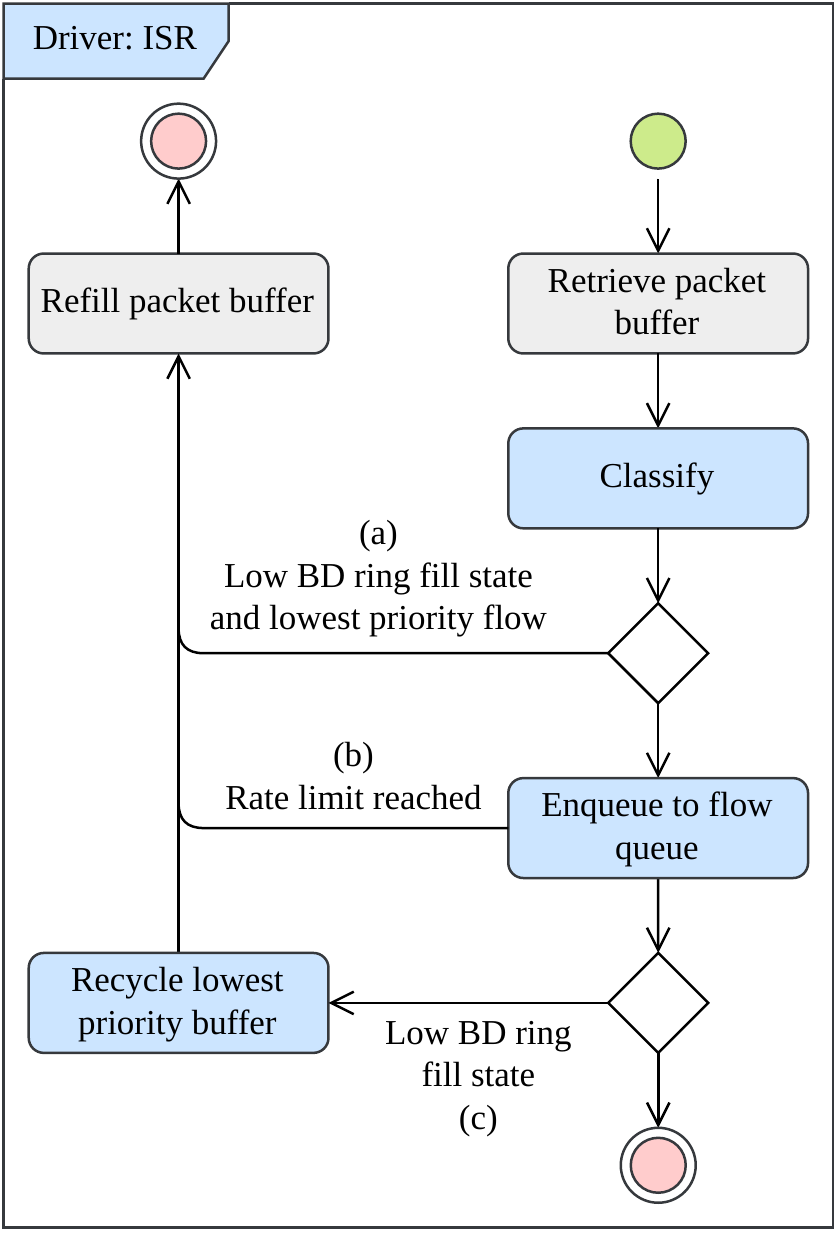}
    \caption{\textbf{Top half driver ISR:} Key execution paths that determine whether and when a packet is dropped to save execution time in high-load scenarios.}
    \label{fig:top_half_activity}
\end{figure}

The resulting activity in the top half driver is depicted in Figure \ref{fig:top_half_activity}.
Notable are the three different execution paths that might be taken:
If due to a low \acs{bd} ring fill state a packet has to be recycled, and the currently considered one is of lowest priority, it gets recycled in a short-circuiting branch \textsuperscript{(a)}.
A flow queue may decline further packets to prevent overload by this particular flow \textsuperscript{(b)}.
Lastly, if the short-circuit branch was not taken but the \acs{bd} ring fill state is low, another packet has to be recycled and inserted into the \acs{bd} ring \textsuperscript{(c)}.

\subsection{Prioritized Protocol Handling}
\label{sec:prioritedprotocolhandling}

Once the heterogeneous incoming packets are demultiplexed into differentiated flow queues, the protocol processing can be carried out with the receiver's priority, as proposed in \cite{druschel1996lazy,lee_interrupt_2010}.

We apply a priority inheritance scheme to the single protocol processing task \cite{mercer1991evaluation}.
It allows the task scheduler to preempt the packet processing at any time. Additionally it keeps a low footprint in terms of task resources as well as integrating nicely into embedded network stacks that commonly use a single network task.

To implement this scheme, the priority of the network task has to be moderated depending on the currently processed packet and waiting packets, in order to avoid priority inversion.
For the priority of the network task it must prevail:
\[ \arraycolsep=1.4pt
p_\text{IP-task} = \max \left( \begin{array}{ll}
     & \{ p(f) \mid f \in \textit{Flows} \land \text{nonempty}(f) \} \\
     \cup & \{ p(f) \mid f \in \textit{Flows} \land f \text{ is processed} \} \\
\end{array} \right) \]
This equation can be satisfied through reconsidering the network task's priority every time a packet gets queued or a packet has been processed.
On packet reception, the priority needs to be elevated \textit{iff} the respective flow priority is higher than the current priority assigned to the network task.
On finished packet processing, the priority needs to be decreased \textit{iff} the highest priority packet waiting in the differentiated flow queues is lower than the current network task priority.
This operation is supported by the ability of the differentiated flow queue data structure to efficiently provide the highest enqueued priority (reconsider Figure~\ref{fig:pdest_depq}).

It may appear that by using priority inheritance carried out per packet, we put an overly high computational burden on the fixed-priority task scheduler.
Other designs could use multiple processing tasks with constant priority, to which packets are assigned according to their flow.
The unblocking operation triggered when the first packet of a particular priority is enqueued then adds at least the same overhead: Some task has to be moved into the priority-respective ready task list, and moved out once blocked again.
Among all possible designs that properly communicate the current packet processing priority demand to the task scheduler, our priority-inheriting one has therefore the lowest possible scheduler data structure manipulation overhead.

In order to also have the deferrable parts of the driver processing scheduled according to packet flows, the networking task dequeues a highest priority packet buffer from the differentiated flow queues and executes the second half of the driver before continuing with the regular processing procedure.

\subsection{Rate Limitation}

To take advantage of Early Demuxing while at the same time keeping the system protected from overload conditions, deterministic mitigation techniques \cite{packetoverloadmitigation} are applied to all but the low priority best-effort flows.
Additionally, the unconditionally executed \ac{isr} that demultiplexes incoming packets could incur a high load even if the subsequent scheduling cuts off further processing.
Hence, an additional, global rate limitation needs to be present.

To apply the rate limitation, each flow is scheduled by a conceptual aperiodic server with each incoming packet being modelled as an aperiodic request.
In our prototype we use the deferrable server scheme (c.f.~\ref{sec:aperiodic}).
Beyond the server capacity, packets are discarded.
For the individual flow queues, this happens as part of the inserting operation (reconsider Figure~\ref{fig:top_half_activity}), in order to avoid a situation with a paused \ac{hp} flow queue full of packets blocking all other processing.

To enforce a global rate limitation, once the capacity has been reached in a period, the driver processing switches from \ac{isr}-based execution to a polling driver task, staying in this mode until the capacity is not immediately reached at the begin of a period anymore.
When not processing packet receive \acp{irq} issued by the \ac{nic}, the \ac{bd} ring is filled until eventually packets are discarded by the \ac{nic}.

\subsection{Limitations}

The ability to proceed with deferred packet processing after a phase of higher system load depends on the number of available packet buffers.
As in our architecture these buffers have to be prepared for immediate \ac{dma} operation and therefore a constant amount is dedicated to the lower levels of processing, additional memory might be necessary.
However, this issue also arises with hardware demultiplexing and prioritization support, as the \acs{bd} ring then has to buffer traffic bursts too.

\acs{ip} fragmentation cannot be dealt with properly in our architecture.
To demultiplex fragmented packets, their reassembly had to be done in the \acs{isr}, jeopardizing its \ac{wcet}.
This design treats all packet fragments as belonging to a background priority flow.
Yet, \ac{ip} fragmentation is discouraged, as it introduces robustness, reliability and security issues \cite{kent1987fragmentation, gilad2011fragmentation}.

\section{Evaluation}
\label{sec:evaluation}

In this section we present empirical results collected from our prototypical implementation and subsequently discuss the effectiveness of our approach.

\subsection{Test Setup}

We run FreeRTOS together with a modified FreeRTOS+TCP on a Xilinx Zynq-7000 processing system. %
Networking is done through a Gigabit-class ethernet interface controlled by a Marvell 88E1518 \ac{phy} controller. Notable features are \ac{dma} and TX/RX-checksum offloading.
For our measurements, only a single processor core is active.
Even though this system is relatively powerful in terms of raw processing power and CPU design in comparison to typical \ac{iot} hardware, it can be fully occupied by packet processing.

We pursued two different complementary approaches for measuring the effect on system load:

\begin{enumerate}
    \item \emph{Passive:} A background worker carries out CPU-intensive work and monitors its performance.
    \item \emph{Active:} The software is instrumented to indicate notable events, i.e. task switches, \acp{irq}, and packet processing.
\end{enumerate}

The former is suitable for precisely estimating the average load that a particular scenario puts on the CPU.
While the latter introduces some overhead in the range of 1-5\% to the processing and misses some of the \ac{irq} switching, it allows us to evaluate the distribution of processing-induced latency.

\subsection{CPU-Time Saved with Early Demultiplexing}

\begin{figure}
    \centering
    \input{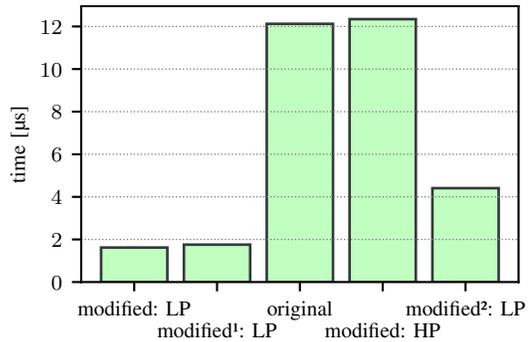}
    \caption{\textbf{Packet processing: }CPU time per zero-length UDP packet under loads between $10^2$ and $10^6$ pkt/s with different configurations.\\
    ¹Short-circuiting branch disabled. ²Cache invalidation deferral disabled.}
    \label{fig:cpu_time_per_packet}
\end{figure}

In this scenario two \ac{udp} sockets are bound, one with a low and one with a high priority receiver process.
To not alter the results, the capacity of all flows as well as the overall \acs{irq} limitation is set to infinity.

Different system configurations were confronted with an endured zero-length \ac{udp}-packet load of constant rate for 60 seconds. Through passive measurement performed by a medium-priority task, the average CPU processing time per packet was then calculated (Figure \ref{fig:cpu_time_per_packet}).
In this experiment we observed that the CPU costs for processing a single packet are rather independent from the magnitude of incoming traffic, staying approximately constant in the range from $10^2$ to $10^6$ pkt/s. %

When \ac{lp} packets get no chance to be scheduled, the executed activity is only that of the Early Demuxing \ac{isr}.
Its average processing duration is $1.62 \mu s$ per packet.
Compared to the original stack as a baseline, which needs $12.1 \mu s$ to fully process a packet, this results in a speedup of 7.5x.
However, due to the short-circuiting logic depicted in Figure \ref{fig:top_half_activity} \textsuperscript{(a)}, in this constant \ac{lp}-flow measurement the packet buffers are discarded without being placed into a flow queue.
When disabling the short-circuiting code path, the per-packet processing time increases to $1.75 \mu s$, still yielding a 7x speedup compared to the full processing in the original stack.

The observed \ac{hp} packets make it through the whole network stack and cause a processing time of $12.3 \mu s$, decreasing receive performance by $1.7$ \% compared to the baseline stack.
This already small relative difference would decrease further if subsequent reception by the receiver task is taken into account, which is obligatory for any soft real-time flow.

If we modify our prototype to again eagerly establish cache coherency in the \acs{isr}, the time spent for \ac{lp} packets increases notably to $4.4 \mu s$.
Therefore, we conclude that incorporating a driver deferral mechanism into our architecture is essential to the performance on cached systems.

\subsection{Packet Processing Latency}

\begin{figure}
    \centering
    \input{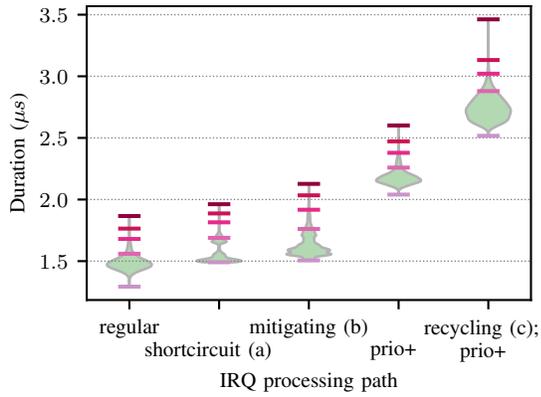}
    \caption{\textbf{ISR processing: }Latency distributions for different execution paths in our modified stack:
    The horizontal bars indicate the percentiles $0\%$, $90\%$, $99\%$, $99.9\%$, $99.99\%$}
    \label{fig:packet_processing_latency}
\end{figure}

The second experiment deals with the predictability of packet processing latencies in the modified IP stack.
Using the active approach, we can reconstruct the precise execution timings of each packet.
Additionally, this allows us to differentiate between the execution paths of the modified driver.
Since the first experiment already yielded precise average timing estimations, the latency deviation can be restricted to relative difference measurements.

The system was flooded with $10^5$ zero-length \ac{udp} packets of two different priorities successively.
Figure~\ref{fig:packet_processing_latency} visualizes the distributions of \acs{isr} processing duration for specific processing paths. For each distribution, the quantiles $0\%, 90\%, 99\%, 99.9\%, 99.99\%$ are visualized as vertical lines, in order to estimate a probabilistic \ac{wcet}.

\ac{lp} packets initially take the fastest path ("regular"), where incoming packets are enqueued without any other processing.
Once the \acs{bd} ring has reached a low fill state, packets have to be recycled.
Since the incoming packets are already at the lowest level present in the differentiated flow queues, the short-circuiting path ("shortcircuit", \textsuperscript{(a)} in Figure \ref{fig:top_half_activity}) is taken.

\ac{hp} packets in contrast can cause a noticeable increase in \acs{isr} processing time.
At each occurrence of such a packet, the network task's priority has to be increased in order to be scheduled subsequently ("prio+").
In case the \acs{bd} ring is already drained by previously received \ac{lp} packets now waiting inside their flow queue, a revocation is needed, adding further processing time ("recycling; prio+").
We also investigated on \ac{hp} packets that get rejected from their flow queue ("mitigating"), yet they behave similarly as shortcircuited packets.

The results show that the first three execution paths are similarly fast, while the ones that include an increase in priority or recycling operations are more costly.
As we discussed in section \ref{sec:prioritedprotocolhandling}, a priority increase can only happen if the flow priority of a received packet is higher than the one of all the currently enqueued ones.
Without the network task being active to process packets and lower the highest enqueued priority again, this is only possible once for each flow in a cascade of increasingly prioritized flows.
Thus, when the system is flooded for some time and \ac{lp} packets start building up in their queues, eventually the faster paths of the \ac{isr} will be taken.

\subsection{Mitigation and Prioritization}

\begin{figure}[b]
    \centering
    \resizebox{\columnwidth}{!}{
    \input{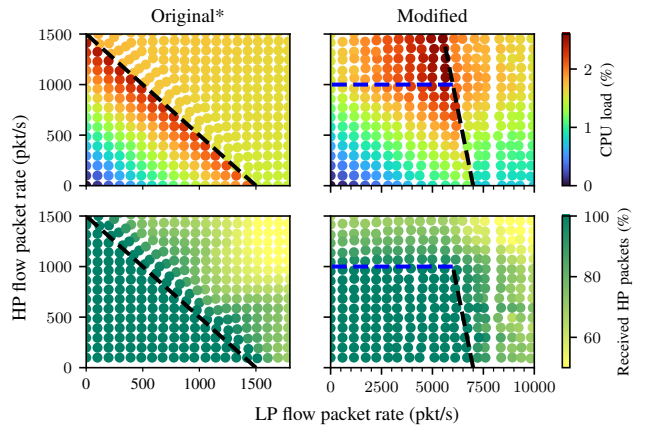}
    }
    \caption{\textbf{Results: }CPU utilization and HP flow liveness at various packet rates on our modified system versus the original system employing only an overall rate limitation. \\
    The blue and black lines mark flow-specific and overall rate limitations respectively.}
    \label{fig:mitigation_map}
\end{figure}

The final results show the effect of protocol processing prioritization and rate limitation, applied both for an individual flow and globally.
Experiments were conducted for multiple combinations of packet flood rates for a \ac{hp}- and \ac{lp}-flow, respectively, over a duration of 3 seconds each.
Again, a medium prioritized task measured the CPU load passively, preventing the scheduling of \ac{lp} packets.
Additionally, a receiver task was employed for the \ac{hp} flow in order to count the packets that arrived at their destination.
Figure \ref{fig:mitigation_map} shows the CPU utilization and the ratio of successfully received \ac{hp} packets to sent ones, as a function of both packet rates.

The original stack was slightly modified to feature an overall \ac{isr} rate limitation, in order to allow a meaningful comparison to our approach.
It is implemented by switching to polling mode once the capacity is reached for a certain period, similar to the one employed in our prototype. In this experiment, the limitations is set at 3 packets per 2 milliseconds.

The CPU utilization increases linearly along with both packet rates, until the global limit of 1500 pkt/s is reached.
Once the polling mode is active, the CPU load drops noticeably.
This can be accounted to the performance improvements gained by switching to a polling-based retrieving activity that handles multiple packets at once.
Further increasing the packet rate causes more \ac{hp} packets to be discarded by the \ac{nic}.

For our modified stack, we chose parameter values anticipating a similar worst case CPU utilization.
Therefore, we configured a high priority flow to allow one packet per millisecond and an unbounded low priority flow.
The \ac{isr} was limited to processing 7000 packets per second.

The CPU load also increases linearly with both packet rates.
As we would expect from the results of the first experiment, the load increases much slower with increasing \ac{lp} packet rates\footnote{notice denser scale}.
Above 1000 pkt/s\footnote{blue line} of \ac{hp} packets, the utilization stagnates as processing of further packets is cut off by the flow queue.
The additional triggered \ac{isr} executions are negligible at this scale.
When the sum of both rates exceeds 7000 pkt/s\footnote{black line}, the CPU utilization also drops with polling activated.
Regarding the liveness of the \ac{hp} flow, we can see how it continuously decreases above the flow-specific rate of 1000 pkt/s.
Additionally, the global limitation impacts the \ac{hp} flow.
So, independent of the \ac{hp} flow rate itself, the communication liveness drops as the system is flooded with \ac{lp} packets.

When comparing our approach to a simple mitigating stack as a baseline, it gets clear that our approach cannot help with processing higher rates of important packets.
Instead, supported by fast Early Demultiplexing and individual prioritization against the remaining tasks, it allows to postpone an overall limitation.
Then, a system can sustain a much higher load of less important packets before it has to go (partially) offline.

\section{Conclusion}
\label{sec:conclusion}

For networked embedded systems running a \ac{rtos}, we present an IP stack design to individually schedule packet processing for differently prioritized \ac{ip}-flows after early demultiplexing.
The issue of costly processing in the network driver is approached by integrating the possibility of deferred buffer processing into our architecture.
Existing embedded IP stacks such as FreeRTOS+TCP and lwIP can be adapted to the proposed design requiring minor modifications.

On our test system, even when having to deal with packet buffers travelling CPU-caches, the CPU load caused by \ac{lp} packets in an already occupied system is reduced by $86\%$ (7x~speedup).
Through limitation parameters, our approach allows system designers to anticipate packet rates of certain soft real-time flows, including those not belonging to any single receiver task, and derive an estimation for the respective processing \ac{wcet}.
Compared to a simple overall mitigation, the design can provide better isolation of processing time allocations for each flow, keeping important flows connected even when other flows exceed their rate limitation.
Budgeting the same CPU resources to the processing of incoming packets, the networking subsystem can still process packets of a \ac{hp}-flow for up to $600\%$ higher overall traffic loads.

Future work could investigate the benefits of using more sophisticated aperiodic scheduling schemes that make use of specific flow's slack time or let flow queues span multiple priority levels, to allow tighter bounds and forgive temporary capacity overruns.

\bibliographystyle{plain}
\bibliography{refs}

\begin{thebibliography}{10}

\bibitem{alcacer2019scanning}
Vitor Alc{\'a}cer and Virgilio Cruz-Machado.
\newblock Scanning the industry 4.0: A literature review on technologies for
  manufacturing systems.
\newblock {\em Engineering science and technology, an international journal},
  22(3), 2019.

\bibitem{amiri2015predictable}
Javad~Ebrahimian Amiri and Mehdi Kargahi.
\newblock A predictable interrupt management policy for real-time operating
  systems.
\newblock In {\em CSI Symposium on Real-Time and Embedded Systems and
  Technologies (RTEST)}. IEEE, 2015.

\bibitem{behnke_interrupting_2021}
Ilja Behnke, Lukas Pirl, Lauritz Thamsen, Robert Danicki, Andreas Polze, and
  Odej Kao.
\newblock Interrupting real-time iot tasks: How bad can it be to connect your
  critical embedded system to the internet?
\newblock In {\em 39th International Performance Computing and Communications
  Conference (IPCCC)}. IEEE, 2020.

\bibitem{behnke2022priority}
Ilja Behnke, Philipp Wiesner, Robert Danicki, and Lauritz Thamsen.
\newblock A priority-aware multiqueue nic design for real-time iot devices.
\newblock In {\em Proceedings of the 35th Annual ACM Symposium on Applied
  Computing (SAC)}. ACM, 2022.

\bibitem{bender2021pieres}
Franz Bender, Jan~Jonas Brune, Nick~Lauritz Keutel, Ilja Behnke, and Lauritz
  Thamsen.
\newblock Pieres: A playground for network interrupt experiments on real-time
  embedded systems in the iot.
\newblock In {\em Companion of the ACM/SPEC International Conference on
  Performance Engineering}. ACM, 2021.

\bibitem{bruckner_introduction_2019}
Dietmar Bruckner, Marius-Petru Stanica, Richard Blair, Sebastian Schriegel,
  Stephan Kehrer, Maik Seewald, and Thilo Sauter.
\newblock An {Introduction} to {OPC} {UA} {TSN} for {Industrial}
  {Communication} {Systems}.
\newblock {\em Proceedings of the IEEE}, 107(6), 2019.

\bibitem{packetoverloadmitigation}
Robert Danicki, Martin Haug, Ilja Behnke, Laurenz M\"{a}dje, and Lauritz
  Thamsen.
\newblock Detecting and mitigating network packet overloads on real-time
  devices in iot systems.
\newblock In {\em Proceedings of the 4th International Workshop on Edge
  Systems, Analytics and Networking (EdgeSys)}. ACM, 2021.

\bibitem{druschel1996lazy}
Peter Druschel and Gaurav Banga.
\newblock Lazy receiver processing (lrp): a network subsystem architecture for
  server systems.
\newblock {\em ACM SIGOPS Operating Systems Review}, 30(SI), 1996.

\bibitem{lwip}
Adam Dunkels.
\newblock Design and implementation of the {lwIP TCP/IP} stack.
\newblock {\em Swedish Institute of Computer Science}, 2(77), 2001.

\bibitem{8412458}
Norman Finn.
\newblock Introduction to time-sensitive networking.
\newblock {\em IEEE Communications Standards Magazine}, 2(2), 2018.

\bibitem{gilad2011fragmentation}
Yossi Gilad and Amir Herzberg.
\newblock Fragmentation considered vulnerable: blindly intercepting and
  discarding fragments.
\newblock In {\em Proceedings of the 5th USENIX conference on Offensive
  technologies}. USENIX, 2011.

\bibitem{gomes_task-aware_2015}
T.~Gomes, P.~Garcia, F.~Salgado, J.~Monteiro, M.~Ekpanyapong, and A.~Tavares.
\newblock Task-aware interrupt controller: Priority space unification in
  real-time systems.
\newblock {\em IEEE Embedded Systems Letters}, 7(1), 2015.

\bibitem{honda2014userlevelstacks}
Michio Honda, Felipe Huici, Costin Raiciu, Joao Araujo, and Luigi Rizzo.
\newblock Rekindling network protocol innovation with user-level stacks.
\newblock {\em ACM SIGCOMM Computer Communication Review}, 44(2), 2014.

\bibitem{humphries2019mind}
Jack~Tigar Humphries, Kostis Kaffes, David Mazi{\`e}res, and Christos
  Kozyrakis.
\newblock Mind the gap: A case for informed request scheduling at the nic.
\newblock In {\em Proceedings of the 18th ACM Workshop on Hot Topics in
  Networks}. ACM, 2019.

\bibitem{jazdi2014cyber}
Nasser Jazdi.
\newblock Cyber physical systems in the context of industry 4.0.
\newblock In {\em International conference on automation, quality and testing,
  robotics}. IEEE, 2014.

\bibitem{smartnics}
Georgios~P. Katsikas, Tom Barbette, Marco Chiesa, Dejan Kosti{\'{c}}, and
  Gerald~Q. Maguire.
\newblock What you need to know about (smart) network interface cards.
\newblock In Oliver Hohlfeld, Andra Lutu, and Dave Levin, editors, {\em Passive
  and Active Measurement}. Springer International Publishing, 2021.

\bibitem{kent1987fragmentation}
C.~A. Kent and J.~C. Mogul.
\newblock Fragmentation considered harmful.
\newblock {\em SIGCOMM Comput. Commun. Rev.}, 17(5), 1987.

\bibitem{realtimehypervisormobileiot}
Neil Klingensmith and Suman Banerjee.
\newblock Hermes: A real time hypervisor for mobile and iot systems.
\newblock In {\em Proceedings of the 19th International Workshop on Mobile
  Computing Systems \& Applications}. ACM, 2018.

\bibitem{lee_priority-based_2015}
Minsub Lee, Hyosu Kim, and Insik Shin.
\newblock Priority-based network interrupt scheduling for predictable real-time
  support.
\newblock {\em Journal of Computing Science and Engineering}, 9(2), 2015.

\bibitem{lee_interrupt_2010}
Minsub Lee, Juyoung Lee, Andrii Shyshkalov, Jaevaek Seo, Intaek Hong, and Insik
  Shin.
\newblock On interrupt scheduling based on process priority for predictable
  real-time behavior.
\newblock {\em {ACM} {SIGBED} Review}, 7, 2010.

\bibitem{mercer1991evaluation}
Clifford~W. Mercer and Hideyuki Tokuda.
\newblock An evaluation of priority consistency in protocol architectures.
\newblock In {\em 16th Conference on Local Computer Networks}, 1991.

\bibitem{niedermaier_you_2018}
Matthias Niedermaier, Jan-Ole Malchow, Florian Fischer, Daniel Marzin, Dominik
  Merli, Volker Roth, and Alexander von Bodisco.
\newblock You snooze, you lose: Measuring {PLC} cycle times under attacks.
\newblock In {\em 12th Workshop on Offensive Technologies (WOOT)}. USENIX,
  2018.

\bibitem{schoeberl_tpip_2018}
Martin Schoeberl and Rasmus~Ulslev Pedersen.
\newblock {tpIP}: {A} {Time}-{Predictable} {TCP}/{IP} {Stack} for
  {Cyber}-{Physical} {Systems}.
\newblock In {\em 21st {International} {Symposium} on {Real}-{Time}
  {Distributed} {Computing} ({ISORC})}. IEEE, 2018.

\bibitem{nics}
Pravin Shinde, Antoine Kaufmann, Timothy Roscoe, and Stefan Kaestle.
\newblock We need to talk about nics.
\newblock In {\em 14th Workshop on Hot Topics in Operating Systems (HotOS)}.
  ACM, 2013.

\bibitem{sprunt1989aperiodic}
Brinkley Sprunt, Lui Sha, and John Lehoczky.
\newblock Aperiodic task scheduling for hard-real-time systems.
\newblock {\em Real-Time Systems}, 1(1), 1989.

\bibitem{strosnider1995deferrable}
Jay~K. Strosnider, John~P. Lehoczky, and Lui Sha.
\newblock The deferrable server algorithm for enhanced aperiodic responsiveness
  in hard real-time environments.
\newblock {\em IEEE Transactions on Computers}, 44(1), 1995.

\end{thebibliography}

\end{document}